\documentclass[12pt, a4paper, twoside]{article}
\usepackage[a4paper, left=3cm,right=2cm]{geometry}
\usepackage{hyperref}
\usepackage{amsfonts}
\usepackage{amsmath}
\usepackage{setspace}
\usepackage{color}

\usepackage[utf8,applemac]{inputenc}
\usepackage{tensor}
\usepackage{cite}
\usepackage{tikz}
\usepackage{graphicx}% Include figure files
\usepackage[font=small, font+=it, format=hang, margin={1cm, 1cm}]{caption}
\graphicspath{{figure/}}
\usepackage{dcolumn}% Align table columns on decimal point
\usepackage{bm}% bold math

\newcommand{\bea}{\begin{eqnarray}}
\newcommand{\eea}{\end{eqnarray}}
\newcommand{\be}{\begin{equation}}
\newcommand{\ee}{\end{equation}}
\def\nn{\nonumber}
\newcommand{\nnr}{\nonumber \\}

\def\p{\partial}
\def\eps{\epsilon}
\newcommand{\R}{\mathbb{R}}
\newcommand{\cR}{\mathcal{R}}
\newcommand{\wt}{\widetilde}
\newcommand{\cM}{\mathcal{M}}

\newcommand{\cV}{\mathcal{V}}
\newcommand{\cF}{\mathcal{F}}
\newcommand{\cI}{\mathcal{I}}
\newcommand{\cN}{\mathcal{N}}
\newcommand{\vph}{\varphi}
\newcommand{\Imag}{\textrm{Im} \,}
\newcommand{\Real}{\textrm{Re} \,}
\newcommand{\Tr}{\textrm{Tr}}
\newcommand{\tr}{\textrm{Tr}}
\newcommand{\wtd}{\widetilde}
\newcommand{\wh}{\widehat}
\newcommand{\im}{\textrm{i}}
   
\title{Non-extremal entropy in $N=2$ supergravities}

\begin{document}

%\maketitle
%\abstract{[...]}

\setcounter{tocdepth}{2}

\begin{titlepage}

\begin{flushright}\vspace{-3cm}
{\small
%{\tt arXiv:yymm.nnnn} \\
\today }\end{flushright}
\vspace{0.5cm}

\begin{center}
{{ \LARGE{\bf{$E_{7(7)}$ invariant non-extremal entropy\\ }}}} \vspace{5mm}

\centerline{\large{\bf{G. Comp\`{e}re\footnote{e-mail: gcompere@ulb.ac.be}, V. Lekeu\footnote{e-mail:
vlekeu@ulb.ac.be}}}}

\vspace{2mm}
\normalsize
\bigskip\medskip
\textit{Universit\'{e} Libre de Bruxelles and International Solvay Institutes\\
CP 231, B-1050 Brussels, Belgium
}
%\vfil
%\pacs{04.70.Dy}

\vspace{25mm}

\begin{abstract}
\noindent 
{The entropy of generic non-extremal dyonic black holes in the STU model has been shown to admit a remarkably universal form. The missing invariant in the formula was recently identified by S\'arosi using the formalism of quantum entanglement as well as a higher dimensional embedding of the U-duality group. Here, we express the non-extremal black hole entropy in the STU model in terms of U-duality covariant tensors. We then provide the extension to the most general non-extremal black hole of ungauged $\mathcal N= 8$ supergravity using $E_{7(7)}$ invariants. We also conjecture a generalization for ungauged $\cN = 2$ supergravity coupled to vector multiplets with arbitrary  cubic prepotential. The most general rotating dyonic black hole solution of the STU model with all scalar moduli turned on is provided in an appendix. }

\end{abstract}

%\pacs{04.65.+e,04.70.-s,11.30.-j,12.10.-g}

\end{center}
%%%%%%%%%%%%%%%%%%%%%%%%%%%%%%%%%%%%%%%%%%%%%%%%%%%%%%%%%%%%%%%%%%%%%%%%%%%%%%%%%%%%%%%%

\end{titlepage}

\newpage
\tableofcontents

\section{Introduction and summary}

Resolving the black hole microstates which underlie the black hole entropy is a major challenge of theoretical physics and of string theory in particular. So far, only supersymmetric black holes have been microscopically counted \cite{Strominger:1996sh,Maldacena:1997de}. Such counting takes place outside of the gravitational regime. No counting is available directly in the gravitational regime, which would otherwise allow to resolve the black hole singularity and help detail the properties of the horizon. We are therefore far from understanding black holes microscopically, especially out of the supersymmetric regime. 

It has been argued that the four-dimensional ungauged $\mathcal N = 8 $ supergravity is the simplest theory of quantum gravity \cite{ArkaniHamed:2008gz} due notably to enhanced symmetries \cite{Cremmer:1978ds} and loop cancellations \cite{Bern:2009kd}. A natural place to investigate the microscopic physics of non-extremal black holes is therefore in such a constrained theory where there is much more control than for the poorly quantum-controlled Schwarzschild or Kerr black hole of Einstein gravity.  More generally, black holes in $\mathcal N = 2$ supergravity already enjoy some enhanced symmetries and loop cancellations and also deserve close study. Much work has been concentrated so far on extremal BPS or non-BPS black holes. In this paper we take one small step in the characterization of black holes away from extremality by writing down the entropy formula for the most general black hole of the STU model, a $\mathcal N= 2$ supergravity with cubic prepotential, its extension to $\mathcal N= 8$ supergravity and other $\mathcal N=2$ supergravities with cubic prepotential.

It has been suggested some time ago that non-extremal black holes are in correspondence with a typical state of strings and D-branes with the same charges of the black hole, with left and right moving excitations \cite{Susskind:1993ws,Horowitz:1996nw}. In its most coarsed grained description, the black hole can be modelled by a single effective string. Its entropy then takes the form 
\bea
S= 2\pi (\sqrt{N_L} + \sqrt{N_R})\label{NLR}
\eea
where $N_{L,R}$ are the left and right moving excitation levels. While it is hard to make this correspondence precise, it is puzzling that the entropy of the most general black hole of the STU model precisely takes the form \eqref{NLR} with   \cite{Chow:2013tia}
\bea
N_L &=& F(M,Q_I,P^I,z_i^\infty) + \Delta(Q_I,P^I) ,\\
N_R &=&  F(M,Q_I,P^I,z_i^\infty)  - J^2.
\eea
Here, most ingredients are well understood: $Q_I,P^I$ are the electromagnetic charges, $J$ is the angular momentum, $M$ the mass and $z_i^\infty$ are the moduli at infinity. The quantity $\Delta(Q_I,P^I)$ is the Cartan quartic invariant which only depends upon the electromagnetic charges. The quantity $F$ is another invariant under U-dualities which was left unidentified in \cite{Chow:2013tia,Chow:2014cca}. Its value is however well-understood at extremality for both branches of regular extremal black holes. The fast rotating extremal black holes admit $F=J^2$ and the slow rotating extremal black holes admit $F = -\Delta$. The entropy then reduces to the two familiar expressions which are only functions of the quartic invariant and the angular momentum. Away from extremality, the attractor mechanism does not apply and the entropy becomes a function of the moduli which is captured by the F-invariant. 

In a recent paper \cite{Sarosi:2015nja}, S\'arosi was able to conveniently write the F-invariant as a polynomial expression in terms of triality invariants of the STU model. In general, the F-invariant is not a polynomial in terms of the charges at infinity, as illustrated by explicit subcases \cite{Chow:2014cca}. The strategy used by S\'arosi was instead to write the invariant in terms of scalar charges, which appear as the first subleading term in the asymptotic expansion of the scalar fields at radial infinity (the leading order being the moduli). The complexity of the invariant is therefore reduced thanks to these auxiliary variables which absorb most of the algebraic complexity. We  checked independently that the manifest triality invariant  formula proposed in \cite{Sarosi:2015nja} for the STU model matches with the entropy formula written in parametric form originally derived in \cite{Chow:2013tia}. 
In \cite{Sarosi:2015nja}, the F-invariant was  however written  in terms of tensors covariant with respect to the $SL(6,\mathbb R)$ embedding of the triality group. Here, we will rewrite this invariant in terms of tensors covariant under triality without any additional embedding. We will then use this invariant to derive the entropy of the generic black hole of $\mathcal N = 8$ supergravity.

The STU model is a consistent truncation of $\cN=8$ supergravity \cite{Cremmer:1984hj,Duff:1995sm}, as reviewed in \cite{Chow:2014cca}. In fact, the generic non-extremal black hole of the STU model can be used as a seed for the generic non-extremal black hole of $\mathcal N = 8$ supergravity upon acting with U-dualities \cite{Sen:1994eb,Cvetic:1996zq}.  Therefore, once the black hole entropy is derived in a triality invariant form for the STU model, one only needs to understand how to generalize these invariants to $E_{7(7)}$ invariants. A proposal for such a map was provided in \cite{Sarosi:2015nja}. This proposal rests on the existence of a matrix $\mathcal R$ transforming under the adjoint of $E_{7(7)}$ such that when restricted to the STU duality frame, it obeys $\tr(\mathcal R^2) = \tr (\tilde R^2)$ and $\tr(\mathcal R^4) = \tr (\tilde R^4)$ where $\tilde R$ is the corresponding matrix of the STU model. In the STU model, one readily obtains that $\tr (\tilde R^2)$ and $\tr (\tilde R^4)$ are independent invariants. Now, this is not the case for $E_{7(7)}$: mathematicians have shown \cite{Lee:1974:IPW,Berdjis:1981:CCC,Berdjis:1981:CO} that $\tr(\mathcal R^2)$ and $\tr(\mathcal R^4)$ are not independent invariants, as we will also cross-check in \eqref{rel}. Therefore, the generalization proposed in \cite{Sarosi:2015nja} lacks one independent invariant and the proposed $E_{7(7)}$ invariant formula cannot reduce to the invariant formula for the STU model.  
In this paper we derive the $E_{7(7)}$ F-invariant formula from first principles using the explicit embedding of the STU model in $\mathcal N = 8$ supergravity, and in accordance with the representation theory for $E_{7(7)}$ \cite{Lee:1974:IPW,Berdjis:1981:CCC,Berdjis:1981:CO}.

Finally, it is interesting to generalize the non-extremal entropy formula to more generic $\mathcal N = 2$ supergravities. In this paper, we will rewrite the invariants in terms of the prepotential formalism in the case of the STU model. This will allow us to propose a conjecture for the non-extremal black hole entropy of any ungauged  $\mathcal N = 2$ supergravity without hypermultiplets with cubic prepotential. We will finally comment on the simpler invariant formula for quadratic prepotentials. 

Our paper is organized as follows. We start by deriving the F-invariant in the STU model in Section \ref{sec:STU}. After a review of the $E_{7(7)}$ algebra and the embedding of the STU model in $\mathcal N = 8$ supergravity we derive the generalization of the F-invariant to $\mathcal N = 8$ supergravity in Section \ref{sec:E7}. Finally, we rewrite the invariant of the STU model in the prepotential formalism in Section \ref{sec:N2} and conjecture that the formula holds for more generic cubic prepotentials.  Appendix \ref{app1} describes the most general rotating dyonic black hole solution to the STU model with all scalar moduli turned on. Appendix \ref{app:so8} contains some conventions.

\section{STU model and triality}
\label{sec:STU}

We will follow the conventions of \cite{Chow:2014cca}. The STU model is the four-dimensional ungauged $\mathcal N=2$ supergravity coupled to 3 vector multiplets with cubic prepotential $F=-X^1X^2X^3/X^0$. This model occurs as a consistent truncation of various supergravity theories. The bosonic content of the action is the metric, four gauge fields $A_\mu^I$, $I=1,2,3,4$, and three complex scalar fields $z_i = x_i + i y_i=\chi_i+ie^{-\varphi_i}$, $i=1,2,3$.   
 The U-duality group of this supergravity is $SL(2,\mathbb R)^3$ together with the finite group of permutations of three elements. This group is usually called the triality U-duality group or triality group in short.

In order to define U-duality invariants that are relevant for describing the F-invariant the first step is to write down covariant tensors in terms of the electromagnetic charges, scalar moduli and scalar charges which transform naturally under the U-duality group. We will do so in Section \ref{covtSTU}. We then describe the construction of invariants in Section \ref{consSTU}, compare them with those of \cite{Sarosi:2015nja} in Section \ref{SL6STU} and conclude with the result for the F-invariant in Section \ref{FinvSTU}.

\subsection{Covariant tensors}
\label{covtSTU}

The electromagnetic charges are conveniently organized as the \emph{charge tensor} $\gamma_{aa'a''}$, with components\footnote{The charge tensor defined in (6.14) in \cite{Chow:2014cca} was incorrect since it does not transform covariantly under $SL(2,\mathbb R)^3$ transformations. With the present correction, no other formula of \cite{Chow:2014cca} is affected.}
\begin{align}
(\gamma_{000}, \gamma_{111}) & = (P^4,-Q_4) , & (\gamma_{100},\gamma_{011}) & = (Q_1,-P^1) , \nnr
(\gamma_{010}, \gamma_{101}) & = (Q_2,-P^2) , & (\gamma_{001}, \gamma_{110}) & = (Q_3,-P^3) ,
\end{align}
where $Q_I$, $P^I$ are the electromagnetic charges. If the reader is more familiar with the symplectic formalism, note that $Q_0=-P^4$, $P^0 = Q_4$. 
The charge tensor transforms as
\be \label{SL2charges}
\gamma_{aa'a''} \mapsto (S_1)\indices{_a^b}(S_2)\indices{_{a'}^{b'}}(S_3)\indices{_{a''}^{b''}}\gamma_{bb'b''}
\ee
under $SL(2,\R)^3$, where the group elements $S_i \in SL(2, \R)_i$ are
\begin{align}
S_i & = 
\begin{pmatrix}
a_i & b_i \\
c_i & d_i
\end{pmatrix}
 & (\text{with } a_i d_i - b_i c_i & = 1).
\end{align}
The scalar fields parametrize the three coset matrices
\be
\mathcal M_i = \frac{1}{y_i}
\begin{pmatrix}
1 & x_i \\
x_i & x_i^2 + y_i^2
\end{pmatrix}
=
\begin{pmatrix}
e^{\varphi_i} &  \chi_i e^{\varphi_i} \\
 \chi_i e^{\varphi_i} & e^{-\varphi_i} + \chi_i^2 e^{\varphi_i}
\end{pmatrix}.
\ee
Each such matrix is invariant under the $SL(2,\mathbb R)_j$ group with $j \neq i$ while they transform under  $SL(2,\mathbb R)_i$ as
\be
\mathcal M_i \mapsto \omega_i^T \mathcal M_i \omega_i ,
\ee
where $\omega_i \in SL(2, \R)_i$ is given by
\begin{align}
\omega_i & = 
\begin{pmatrix}
d_i & b_i \\
c_i & a_i
\end{pmatrix}
 & (\text{with } a_i d_i - b_i c_i & = 1) .
\end{align}
Notice that $\omega_i \neq S_i$. In fact, defining $\sigma = \begin{pmatrix} 1 & 0 \\ 0 & -1 \end{pmatrix}$, we have
\be \omega_i = \sigma (S_i)^{-1} \sigma .\ee
Since we aim at finding covariant tensors, it is therefore convenient to first define new coset matrices as
\be \wt{\cM}_i = \sigma \cM_i \sigma =\frac{1}{y_i}
\begin{pmatrix}
1 & -x_i \\
-x_i & x_i^2 + y_i^2
\end{pmatrix}\ee
that transform as
\be \wt{\cM}_i \mapsto (S_i^{-1})^T \wt{\cM}_i S_i^{-1}. \ee
Such a coset matrix is known to admit the asymptotic expansion
\be \wt{\cM}_i = \wt{\cM}_i^{(0)} + \frac{\wt{\cM}_i^{(1)}}{r} + \mathcal{O}\left(\frac{1}{r^2}\right). \ee
Here $\wt{\cM}_i^{(0)}$ encodes the scalar moduli at infinity $z_i^\infty$ while $\wt{\cM}_i^{(1)}$ encodes the scalar charges $\Sigma_i$, $\Xi_i$ defined as
\begin{align}
\varphi_i &= \varphi_i^{\infty} + \frac{\Sigma_i}{r} + \mathcal{O}(r^{-2}), & \chi_i &= \chi_i^{\infty} + \frac{\Xi_i}{r} + \mathcal{O}(r^{-2}).
\end{align}
We define the \emph{dressed scalar charge tensor} as
\be R_i = (\wt{\cM}_i^{(0)})^{-1} \wt{\cM}_i^{(1)} .\label{ds}\ee
This tensor is invariant under $SL(2,\mathbb R)_j$ with $j \neq i$ and transforms under $SL(2,\mathbb R)_i$ as
\be R_i \mapsto S_i R_i S_i^{-1} .\ee
For trivial scalar moduli (asymptotically flat boundary conditions), we have
\begin{align}
\wt{\cM}_i^{(0)} &= \begin{pmatrix} 1 & 0 \\ 0 & 1 \end{pmatrix}, & R_i &= \begin{pmatrix} \Sigma_i & -\Xi_i \\ -\Xi_i & -\Sigma_i \end{pmatrix} .\label{trivscal}
\end{align}
Since the $SL(2,\R)^3$ transformations have a non-trivial action on the scalar fields, U-dualities do not preserve trivial scalar moduli and will therefore not preserve the simple form \eqref{trivscal}.

Finally, we also have the invariant tensor for each copy of $SL(2,\mathbb R)_i$
\be \varepsilon = \begin{pmatrix} 0 & 1 \\ -1 & 0 \end{pmatrix}, \ee
that satisfies
\be \Lambda^T \varepsilon \Lambda = \Lambda \varepsilon \Lambda^T = \varepsilon \ee
for any $\Lambda$ in $SL(2,\R)$.

Let us now summarize our ingredients and their transformation laws under $S_1 \otimes S_2 \otimes S_3 \in SL(2,\R)^3$, in index notation. To avoid notational clutter, we write $\wt{\cM}_i^{(0)}=M_i$ from now on. We have the following objects:
\begin{itemize}
\item charge tensor $\gamma_{aa'a''} \mapsto (S_1)\indices{_a^b}(S_2)\indices{_{a'}^{b'}}(S_3)\indices{_{a''}^{b''}}\gamma_{bb'b''}$;
\item asymptotic coset tensors $(M_i)^{ab} \mapsto (M_i)^{cd} (S_i^{-1})\indices{_c^a}(S_i^{-1})\indices{_d^b}$;
\item dressed scalar charge tensors $(R_i)\indices{_a^b} \mapsto (S_i)\indices{_a^c} (R_i)\indices{_c^d} (S_i^{-1})\indices{_d^b}$.
\end{itemize}
We can also use the invariant epsilon tensor $\varepsilon^{ab}$.

\subsection{Construction of invariants}
\label{consSTU}

To build triality invariants, we proceed in two steps.
\begin{enumerate}
\item First, we make $SL(2,\R)^3$ invariants by contracting all indices, with the constraint that only indices corresponding to the same $SL(2,\R)$ can be contracted together.
\item Second, we implement invariance under permutations of the three $SL(2,\R)$ factors by summing the expression with all others obtained by permuting its different $SL(2,\mathbb R)$ internal indices. The result is then automatically invariant under triality. In general there are 6 terms but symmetries might reduce them to 3 or only 1 term. For example, from the $SL(2,\R)^3$-invariant expression
\be \varepsilon^{ab}\varepsilon^{a'b'}M_3^{a''b''} \gamma_{aa'a''}\gamma_{bb'b''} \ee
we make
\be \left( M_1^{ab}\varepsilon^{a'b'}\varepsilon^{a''b''} + \varepsilon^{ab}M_2^{a'b'}\varepsilon^{a''b''} +\varepsilon^{ab}\varepsilon^{a'b'}M_3^{a''b''}  \right) \gamma_{aa'a''}\gamma_{bb'b''} . \ee
\end{enumerate}
There are only three terms because the permutations of two $\varepsilon^{ab}$ tensors give identical terms. 

We define the degree as follows: the mass, NUT charge, electromagnetic charges and scalar charges have degree 1 while the moduli have degree 0. Therefore, $M_i^{ab}$ and $\varepsilon^{ab}$ have degree 0 while $\gamma_{aa'a''}$ and $(R_i)\indices{_a^b}$ have degree 1. Inspection reveals that the F-invariant is a homogeneous function of degree 4. Restricting to degree $\leq 4$, we find the following independent invariants:
\begin{itemize}
\item Degree 1:
\begin{align}
M, \, N.
\end{align}
\item Degree 2:
\begin{align}
L_1 &= M_1^{ab}M_2^{a'b'}M_3^{a''b''} \gamma_{aa'a''}\gamma_{bb'b''}, \\
L_2 &= \frac{1}{3}  \left( \Tr{R_1^2}+\Tr{R_2^2}+\Tr{R_3^2} \right) .
\end{align}
\item Degree 3:
\begin{align}
C_1 &= \frac{1}{3} \sum \varepsilon^{ac} R\indices{_{1c}^b} \varepsilon^{a'b'}\varepsilon^{a''b''} \gamma_{aa'a''}\gamma_{bb'b''} ,\\
C_2 &= \frac{1}{3} \sum M_1^{ac} R\indices{_{1c}^b} M_2^{a'b'}M_3^{a''b''} \gamma_{aa'a''}\gamma_{bb'b''} .
\end{align}
\item Degree 4:
\begin{align}
\Delta &= \frac{1}{32} \varepsilon^{ac} \varepsilon^{a'b'} \varepsilon^{a''b''} \varepsilon^{bd} \varepsilon^{c'd'} \varepsilon^{c''d''} \gamma_{aa'a''}\gamma_{bb'b''} \gamma_{cc'c''}\gamma_{dd'd''} ,\\
\Delta_2 &= \frac{1}{96} \sum M_1^{ac} \varepsilon^{a'b'} \varepsilon^{a''b''} M_1^{bd} \varepsilon^{c'd'} \varepsilon^{c''d''} \gamma_{aa'a''}\gamma_{bb'b''} \gamma_{cc'c''}\gamma_{dd'd''} ,\\
\Delta_3 &=\frac{1}{96}  \left( \Tr{R_1^4}+\Tr{R_2^4}+\Tr{R_3^4} \right).
\end{align}
\end{itemize}
Here, each sum is over the three cyclic permutations of the $SL(2,\mathbb R)$ indices. The familiar quartic invariant is $\Delta$. Many more invariants can be formulated but we will not classify them here. This list will be sufficient to express the F-invariant below. 

\subsection{Reformulation in a $SL(6,\mathbb R)$ embedding}
\label{SL6STU}

In his recent paper \cite{Sarosi:2015nja}, S\'arosi constructed U-duality invariants using the embedding of the U-duality group into $SL(6,\mathbb R)$. The $SL(2,\mathbb R)^3$ transformations are expressed as $S \in SL(6,\mathbb R)$ with 
\bea
S = \left( \begin{array}{ccc} S_1&&\\ & S_2& \\&& S_3 \end{array}\right)
\eea
while the 6 permutations of the triality group are represented by block permutation matrices. 

One starts from the pair of tensors
\bea
(\psi_1)_{aa'a''} = -\gamma_{aa'a''},\qquad (\psi_2)_{aa'a''} = \widetilde \gamma_{aa'a''},
\eea
where $\widetilde \gamma_{aa'a''}$ is obtained from $\gamma_{aa'a''}$ by electromagnetic duality:
\bea\label{tQP}
\widetilde Q^I  =  P^I ,\qquad \widetilde P_I  = - Q_I.
\eea
We denote them as $(\psi_\alpha)_{aa'a''} $ with $\alpha=1,2$.
One can then construct from $(\psi_\alpha)_{aa'a''}$ an antisymmetric $SL(6,\mathbb R)$-covariant tensor $(P_{\psi_\alpha})_{ABC}$ ($A,B,C=1,\dots, 6$) according to
\be (P_{\psi_\alpha})_{a+1, a'+3, a''+5} = (\psi_\alpha)_{a a' a''} \qquad (a,a',a''=0,1).  \ee
The other components of $P_{\psi_\alpha}$ are either obtained from those by antisymmetry or are zero.
It transforms as
\be
(P_{\psi_\alpha})_{ABC} \mapsto S_A^{\;\; A'}S_B^{\;\; B'}S_C^{\;\; C'} (P_{\psi_\alpha})_{A'B'C'}.
\ee
Finally, one builds the four $SL(6,\mathbb R)$-covariant tensors 
\bea
(K_{\alpha\beta})^A_{\; B} &=&\frac{1}{2!3!} \eps^{A C_1C_2C_3C_4C_5} (P_{\psi_\alpha})_{B C_1C_2}(P_{\psi_\beta})_{C_3C_4C_5}
\eea
which transform as
\be
K_{\alpha\beta} \mapsto (S^{-1})^T K_{\alpha\beta} S^T.
\ee

One can also construct a block-diagonal matrix $R$ which contains the three copies of $R^T_i$ where $R_i$, $i=1,2,3$, is defined in \eqref{ds}. It transforms as 
\bea
R \mapsto (S^{-1})^T R S^T. 
\eea

Restricting to degree $\leq 4$, S\'arosi then found a list of independent invariants, which we relate to the invariants defined in Section \ref{consSTU} as
\begin{itemize}
\item Degree 1:
\begin{align}
M,\, N.
\end{align}
\item Degree 2:
\begin{align}
\Tr(K_{12}) &= -3 L_1 ,\\
\Tr(R^2) &= 3 L_2.
\end{align}
\item Degree 3:
\begin{align}
\Tr(K_{11}R) &= 3 C_1 ,\\
\Tr(K_{12}R) &= - 3 C_2.
\end{align}
\item Degree 4:
\begin{align}
\Tr(K_{11}^2) &= - 96 \Delta, \\
\Tr(K_{11}K_{22}) &= - 96 \Delta_2, \\
\Tr(R^4) &= 96 \Delta_3.\label{TrR4}
\end{align}
\end{itemize}

\subsection{The F-invariant}
\label{FinvSTU}

The area over $4G$ of the outer and inner horizons of the general non-extremal black hole of the STU model (with NUT charge included for completeness) takes the form 
\bea
S_{\pm} =2\pi \left(  \sqrt{F+\Delta} \pm \sqrt{F - J^2}  \right)
\eea
where the $F$ invariant is given in terms of the auxiliary parameters $m,n,\nu_1,\nu_2$ defined in \cite{Chow:2014cca} as
\bea
F = \frac{(m^2+n^2)}{G^2}(-n \nu_1 + m \nu_2)^2.\label{iF}
\eea
In  \cite{Sarosi:2015nja}, this formula was matched after extensive and convincing numerical tests to the following formula written in terms of triality invariants built from the  $SL(6,\mathbb R)$ embedding of the triality group,
\bea
\label{autrelabel}
F&=&M^4+M^2 N^2+\frac{ M^2}{12}\text{Tr}  K_{12}-\frac{M }{24}\text{Tr} ( K_{12}R)+\frac{N^2}{24}\text{Tr}(R^2)\nn \\ 
&&-\frac{N}{24}\text{Tr}(K_{11} R)+\frac{1}{192}\left(\text{Tr}(K_{11}^2)-\text{Tr}(K_{11}K_{22})-\frac{1}{2}(\text{Tr}R^2)^2+\text{Tr}(R^4)\right).
\eea
We checked independently that formulae \eqref{iF} and \eqref{autrelabel} numerically agree for around a hundred random values of the parameters.

Using this and the dictionary above we obtain the formula for the F-invariant in terms of triality invariants which are functions of the mass, the NUT charge, the electromagnetic charges, the moduli and the scalar charges:
\bea\label{Finv1}
F = M^4 +M^2 N^2-\frac{M^2}{4} L_1 +\frac{N^2}{8}L_2 + \frac{M}{8}C_2 -\frac{N}{8}C_1+ \frac{- \Delta +\Delta_2 +\Delta_3}{2}  -\frac{3}{128}(L_2)^2.
\eea
One can obviously set the NUT charge to zero, $N=0$, to find physical configurations. 

It is in principle possible to write the scalar charges in terms of the physical charges (the mass, the electromagnetic charges and the scalar moduli) but this process is algebraically complicated. For the example of the Kaluza-Klein black hole, an explicit formula in terms of the physical charges could be obtained but involved trigonometric functions \cite{Chow:2014cca}. For the four-charge Cveti{c}-Youm black hole \cite{Cvetic:1996kv}, it has been stated  in \cite{Sarosi:2015nja} that one encounters a fifth order polynomial in the inversion algorithm, which therefore cannot be solved by radicals.

\section{$\mathcal N =8$ supergravity and $E_{7(7)}$}
\label{sec:E7}

The STU model is a consistent truncation of ungauged $\cN=8$ supergravity \cite{Cremmer:1984hj,Duff:1995sm}, as reviewed in \cite{Chow:2014cca}. As we already mentioned, the generic non-extremal black hole of the STU model can be used as a seed for the generic non-extremal black hole of $\mathcal N = 8$ supergravity \cite{Sen:1994eb,Cvetic:1996zq}. We derived the black hole entropy for the STU model in terms of triality invariants in the previous section. We will now reformulate the black hole entropy in terms of $E_{7(7)}$ invariants by matching individual invariants in $\mathcal N = 8$ supergravity with their corresponding invariants in the STU U-duality frame.  This will give the correct formula for the most general non-extremal black hole of $\mathcal N = 8$ supergravity.

\subsection{Preliminary: the $\mathfrak{e}_{7(7)}$ algebra}

Let first us summarize some key features of the $\mathfrak{e}_{7(7)}$ algebra. We will follow the conventions of \cite{Cremmer:1979up}.

We parametrize an element of $\mathfrak{e}_{7(7)}$ by a traceless $8\times 8$ matrix $\Lambda\indices{^i_j}$ and an antisymmetric tensor $\Sigma_{ijkl}$ (indices go from $1$ to $8$). It acts in the fundamental representation on a pair of antisymmetric tensors $(X^{ij}, X_{ij})$ as
\begin{align}
\delta X^{ij} &= \Lambda\indices{^i_k} X^{kj} + \Lambda\indices{^j_k} X^{ik} + \star\Sigma^{ijkl}X_{kl} ,\nnr
\delta X_{ij} &= - X_{kj} \Lambda\indices{^k_i} - X_{ik} \Lambda\indices{^k_j} + \Sigma_{ijkl}X^{kl}
\end{align}
where we defined $\star\Sigma^{ijkl}=\frac{1}{4!}\varepsilon^{ijklmnpq}\Sigma_{mnpq}$. (Remark: the two tensors $X^{ij}$ and $X_{ij}$ are different, independent objects.)
In matrix form, this is
\be \label{eq:E7matrices}
\delta \begin{pmatrix} X^{ij} \\ X_{ij} \end{pmatrix} =
\begin{pmatrix}
2\Lambda\indices{^{[i}_{[k}}\delta^{j]}_{l]} & \star\Sigma^{ijkl} \\
\Sigma_{ijkl} & -2\Lambda\indices{^{[i}_{[k}}\delta^{j]}_{l]}
\end{pmatrix}
\begin{pmatrix} X^{kl} \\ X_{kl} \end{pmatrix},
\ee
and this gives an explicit representation of $\mathfrak{e}_{7(7)}$ by $56\times 56$ matrices.
The commutator $[(\Lambda_1,\Sigma_1), (\Lambda_2,\Sigma_2)]=(\Lambda_3,\Sigma_3)$ of two $\mathfrak{e}_{7(7)}$ elements is given by
\begin{align}
\Lambda\indices{_3^i_j} &= \Lambda\indices{_1^i_k}\Lambda\indices{_2^k_j} - \frac{1}{3} \star\Sigma_1^{iklm} \Sigma_{2klmj} - (1\leftrightarrow 2), \\  
\Sigma_{3ijkl} &= 4 \Lambda\indices{_1^m_{[i} }\Sigma_{2jkl]m} - (1\leftrightarrow 2) .
\end{align}

Writing the generators as
\begin{align}
G\indices{_a^b} &= (\Lambda\indices{_a^b},0), & (\Lambda\indices{_a^b})\indices{^i_j}&=\delta^i_a\delta^b_j-\frac{1}{8}\delta^a_b\delta^i_j ,\\
G^{abcd}&=(0,\Sigma^{abcd}), & (\Sigma^{abcd})_{ijkl} &= \delta^{[a}_i\delta^b_j\delta^c_k\delta^{d]}_l,
\end{align}
we get the commutation relations
\begin{align} \label{e7comm}
[ G\indices{_a^b}, G\indices{_c^d}] &= \delta^b_c G\indices{_a^d}-\delta^d_a G\indices{_c^b}, \\
[ G\indices{_a^b}, G^{cdef}] &= 4 \delta^{[c}_a G^{def]b} + \frac{1}{2}\delta^a_b G^{cdef}, \\
[ G^{abcd}, G^{efgh} ] &=\frac{1}{72} \left( G\indices{_k^{[a}} \varepsilon^{bcd]efghk} - G\indices{_k^{[e}}\varepsilon^{fgh]abcdk}\right) .
\end{align}

Two subalgebras of interest of $\mathfrak{e}_{7(7)}$ are
\begin{itemize}
\item The manifest $\mathfrak{sl}(8,\R)$ which is generated by $G_a^{\;\; b}$ alone;
\item The maximal compact subalgebra $\mathfrak{su}(8)$, generated by transformations with $\Lambda\indices{^i_j}=-\Lambda\indices{^j_i}$ and $\Sigma_{ijkl}=-\star\Sigma^{ijkl}$.
\end{itemize}

The maximal compact subalgebra $\mathfrak{su}(8)$ can be made manifest via the change of basis
\be X_{AB} = \frac{1}{4\sqrt 2} \left( X^{ij} +i X_{ij} \right) (\Gamma^{ij})_{AB}, \ee
where the $8\times 8$ matrices $\Gamma^{ij}$ are the $\mathfrak{so}(8)$ generators described in appendix \ref{app:so8}.
In this basis, the $\mathfrak{e}_{7(7)}$ transformations are given by 
\be \delta X_{AB} = \Lambda\indices{_A^C} X_{CB} + \Lambda\indices{_B^C} X_{AC} + \Sigma_{ABCD}\bar{X}^{CD}, \ee
where $\bar{X}^{AB} =(X_{AB})^*$, $\Lambda\indices{_A^B}$ is an element of $\mathfrak{su}(8)$, and $\Sigma_{ABCD}$ is a complex antisymmetric tensor satisfying the self-duality condition
\be \Sigma_{ABCD} = \frac{1}{24} \varepsilon_{ABCDEFGH} \bar{\Sigma}^{EFGH}, \qquad \bar{\Sigma}^{ABCD} = (\Sigma_{ABCD})^* .\ee The link between these parameters and $(\Lambda^i_{\;\; j},\Sigma_{ijkl})$ is given in \cite{Cremmer:1979up}. In this basis, the $\mathfrak{su}(8)$ subalgebra is obtained by simply setting $\Sigma_{ABCD} = 0$, whereas the $\mathfrak{sl}(8,\R)$ subalgebra is more involved.

\subsection{Symplectic and quartic invariants}

We can directly identify two invariants:
\begin{itemize}
\item the symplectic invariant, which is a quadratic form over two distinct fundamental representations
\be \label{eq:symplecticE7} X^T \Omega Y = X^{ij}Y_{ij} - X_{ij}Y^{ij}, \ee
where $\Omega = \begin{pmatrix} 0 & I_{28\times 28} \\ -I_{28\times 28} & 0 \end{pmatrix}$;
\item the quartic invariant, which is a quartic form over one fundamental representation
\begin{align}\label{eq:I4E7}
\mathcal{I}_4 (X) = &X^{ij}X_{jk}X^{kl}X_{li} -\frac{1}{4}(X^{ij}X_{ij})^2 \nnr
&+\frac{1}{96} \varepsilon^{ijklmnpq} X_{ij}X_{kl}X_{mn}X_{pq} +\frac{1}{96} \varepsilon_{ijklmnpq} X^{ij}X^{kl}X^{mn}X^{pq} .
\end{align}
\end{itemize}
Remark: the invariance of the symplectic invariant under $\mathfrak{e}_{7(7)}$ transformations \eqref{eq:E7matrices} proves the embedding $\mathfrak{e}_{7(7)} \subset \mathfrak{sp}(56,\R)$.

Since the transformation laws in the $SL(8,\R)$ and in $SU(8)$ bases are formally identical, we can construct invariants in the $SU(8)$ basis by replacing $i,j,\dots$ indices by $A,B,\dots$ indices in the previous invariants. Therefore, the two quantities
\begin{align}
\diamondsuit(X) = &\bar{X}^{AB}X_{BC}\bar{X}^{CD}X_{DA} -\frac{1}{4}(\bar{X}^{AB}X_{AB})^2 \nnr
&+\frac{1}{96} \varepsilon^{ABCDEFGH} X_{AB}X_{CD}X_{EF}X_{GH} +\frac{1}{96} \varepsilon_{ABCDEFGH} \bar{X}^{AB}\bar{X}^{CD}\bar{X}^{EF}\bar{X}^{GH}, \\
(X,Y)_\Omega =& \bar{X}^{AB}Y_{AB} - X_{AB} \bar{Y}^{AB} 
\end{align}
are $E_{7(7)}$-invariant. In fact, the invariants constructed in the two bases are proportional to each other: we have the relations $\diamondsuit(X)=-\mathcal{I}_4(X)$ \cite{Gunaydin:2000xr} and $X^T \Omega Y  = -i (X,Y)_\Omega$, as we checked.

\subsection{Construction of additional invariants}

The $56$ charges of $\cN=8$ supergravity transform in the fundamental representation of $E_{7(7)}$. Accordingly, we can pack them into the tensor $(X_{ij},X^{ij})$, transforming as
\be \delta X = g X \ee
under $g \in \mathfrak{e}_{7(7)}$ (we use the $56\times 56$ matrix notation here).
The $70$ scalar fields parametrize the coset $E_{7(7)}/SU(8)$ in the Borel gauge and are therefore contained in a matrix $\cV$ that transforms as
\be \delta \cV = k \cV - \cV g \ee under $(k,g) \in \mathfrak{su}(8)_{\text{local}} \times \mathfrak{e}_{7(7)}$. The local transformation $k$ is the compensator required to keep $\cV$ in the Borel gauge. Under the action of the group, these transformations are
\begin{align}
X &\mapsto G X, \\
\cV &\mapsto K\cV G^{-1}.
\end{align}
where $K \in SU(8)$, $G \in E_{7(7)}$. In particular, the object $\cV X$ only transforms under $SU(8)$.
From $\cV$, we define the usual matrix
\be \cM = \cV^T \cV, \ee which transforms as
\be \cM \mapsto (G^{-1})^T \cM G^{-1}. \ee
Again, from the asymptotic expansion
\be \cM = \cM^{(0)} + \frac{\cM^{(1)}}{r} + \mathcal{O}\left(\frac{1}{r^2}\right), \ee
we define the dressed charge matrix
\be \cR = (\cM^{(0)})^{-1} \cM^{(1)} \ee
that transforms in the adjoint representation of $E_{7(7)}$,
\be \cR \mapsto G \cR G^{-1} .\ee
In building invariants, we will also make use of $\Omega$, which has the property
\be G^T \Omega G = \Omega \ee
for any $G$ in $E_{7(7)}\subset Sp(56,\R)$.

We can now construct several additional invariants:
\begin{itemize}
\item Since both $X$ and $\cR X$ transform in the fundamental, we can make the following invariants of order two and three:
\be X^T \cM^{(0)} X, \quad X^T \cM^{(0)} \cR X, \quad X^T \Omega \cR X .\ee
\item As noted above, $\cV X$ transforms only under $SU(8)$. Switching to the $SU(8)$ notation (with indices $A,B,\dots$), we can make invariants simply by contracting indices, e.g.
\begin{align}
T_2 &= (\cV X)_{AB} \overline{(\cV X)}^{BA} ,\\
T_4 &= (\cV X)_{AB} \overline{(\cV X)}^{BC} (\cV X)_{CD} \overline{(\cV X)}^{DA}
\end{align}
where as before $\overline{(\cV X)}^{AB} = ((\cV X)_{AB})^*$. Higher order invariants can also be constructed in the same fashion, but we will not need them here. We have the relation
\be T_2 = - X^T \cM^{(0)} X, \ee
so we discard $T_2$ from our list of invariants.
\item Since $\cR$ transforms in the adjoint, all traces
\be \Tr(\cR^k) \ee
of powers of $\cR$ are invariant. These invariants are not all independent; in fact, those with odd $k$ vanish identically. We checked that $\Tr \cR^2$, $\Tr \cR^6$ and $\Tr \cR^8$ are independent, while for $\Tr \cR^4$ we have the relation
\be 
\Tr \cR^4 = \frac{1}{24} (\Tr \cR^2 )^2 . \label{rel}
\ee
In fact, it is known by mathematicians \cite{Lee:1974:IPW,Berdjis:1981:CCC,Berdjis:1981:CO} that the only independent ones are those with
\bea
k = 2, 6, 8, 10, 12, 14 \text{ and } 18 .
\eea
This will introduce a subtlety in identifying the $E_{7(7)}$ generalization of the triality invariant $\text{Tr}(R^4)=96 \Delta_3$ \eqref{TrR4}: it will be a non-polynomial expression in $\Tr \cR^2$, $\Tr \cR^6$ and $\Tr \cR^8$ as we will describe below.
\end{itemize}

\subsection{Embedding of the STU model}

Using the information given in \cite{Chow:2014cca} as well as the explicit formulae for dimensional reduction given in \cite{Cremmer:1997ct}, we can get the explicit embedding of the STU model in $\cN=8$ supergravity.

In the paper \cite{Cremmer:1997ct}, another parametrization of the Borel subalgebra of $\mathfrak{e}_{7(7)}$ is used. 
The link with our notation is
\be \label{eq:E7dict} E\indices{_i^j} = G\indices{_i^j}, \quad E^{ijk} = - 12 G^{ijk8}, \quad D_i = G\indices{_i^8}, \quad \vec{H} = \sum_{j=1}^7 \left( -\vec{f}_j+\vec{g} \right) G\indices{_j^j}, \ee
where the $7$-component vectors $\vec{f}_j$ and $\vec{g}$ are given by
\begin{align}
\vec{f}_j &= (\underbrace{0, \dotsc, 0}_{j-1}, (10-j) s_j, s_{j+1}, \dotsc, s_7) ,\\
\vec{g} &= 3 (s_1, s_2, \dotsc, s_7) ,\\
s_i &= \sqrt{\frac{2}{(10-i)(9-i)}}.
\end{align}
It can be checked using \eqref{e7comm} that these identifications correctly reproduce the commutation relations of \cite{Cremmer:1997ct}.

In this notation, the embedding is the following.
\begin{itemize}
\item The electromagnetic charges are given by
\begin{align}
(X^{12}, X^{34}, X^{56}, X^{78}) &= (Q_1,Q_2,Q_3,-Q_4) ,\nnr
(X_{12}, X_{34}, X_{56}, X_{78}) &= (P^1,P^2,P^3,-P^4) ,
\end{align}
the other $X^{ij}$, $X_{ij}$ being zero.
\item The coset matrix $\cV$ is
\be 
\cV=\exp\left[\frac{1}{2}\sum_{i=1}^3 \varphi_i \vec{v}_i \cdot \vec{H}\right]\exp\left[-\chi_1 E^{127}-\chi_2 E^{347} -\chi_3 E^{567}\right],
\ee
where
\begin{align}
\vec{v}_1 &= \left(\frac{1}{2},\frac{3}{2 \sqrt{7}},-\frac{1}{2\sqrt{21}},-\frac{1}{2 \sqrt{15}},-\frac{1}{2 \sqrt{10}},-\frac{1}{2 \sqrt{6}},\frac{1}{\sqrt{3}}\right), \nnr
\vec{v}_2 &= \left(-\frac{1}{4},-\frac{3}{4 \sqrt{7}},\frac{2}{\sqrt{21}},\frac{2}{\sqrt{15}},-\frac{1}{2 \sqrt{10}},-\frac{1}{2 \sqrt{6}},\frac{1}{\sqrt{3}}\right), \nnr
\vec{v}_3 &= \left(-\frac{1}{4},-\frac{3}{4 \sqrt{7}},-\frac{\sqrt{3}}{2 \sqrt{7}},-\frac{\sqrt{3}}{2 \sqrt{5}},\frac{1}{\sqrt{10}},\frac{1}{\sqrt{6}},\frac{1}{\sqrt{3}}\right),
\end{align}
and three orthonormal vectors and we used the identification \eqref{eq:E7dict} to get the explicit form of $\cV$ as a $56\times 56$ matrix.
\end{itemize}

We can now compare our $E_{7(7)}$ invariants computed in the STU truncation with those of the previous sections. We find
\begin{itemize}
\item Order two:
\begin{align}
L_1&=  X^T \cM^{(0)} X, \\
L_2 &= \frac{1}{36} \Tr(\cR^2).
\end{align}
\item Order three:
\begin{align}
C_1 &=\frac{1}{3} X^T \Omega \cR X, \\
C_2&= \frac{1}{3} X^T \cM^{(0)} \cR X.
\end{align}
\item Order four:
\begin{align}
\Delta &= - \frac{1}{16} \cI_4(X), \label{I4} \\
\Delta_2 &= \frac{1}{96} \left(  8 T_4 + 6 \cI_4 - (X^T \cM^{(0)} X)^2\right) , \\
0 &= 2^{17} 3^7 5 (\Delta_3 )^2 - 2^8 3^3 5 \Delta_3 (\Tr(\cR^2))^2 - 5 (\Tr(\cR^2))^4 \nnr
&\quad + 2^5 3^2 11 \Tr(\cR^2) \Tr(\cR^6) - 2^6 3^5 \Tr(\cR^8). \label{eq:TrR4}
\end{align}
\end{itemize}
As announced, we find a non-polynomial expression for $\Delta_3$,
\be 
\Delta_3 = \frac{1}{2^{10} 3^4 5} \left[ 5 \Tr(\cR^2)^2 + \sqrt{5} \sqrt{5^3 \Tr(\cR^2)^4 - 2^8 3^3 11 \Tr(\cR^2) \Tr(\cR^6) + 2^9 3^6 \Tr(\cR^8)} \right].
\ee
We checked that only this root of \eqref{eq:TrR4} correctly reproduces $\Tr(R^4)$.

\subsection{The F-invariant}

Using formula \eqref{Finv1} and the dictionary above, we find the following formula for the F-invariant in terms of $E_{7(7)}$ invariants:
\begin{align}
F = &M^4+M^2 N^2 - \frac{M^2}{4}X^T \cM^{(0)} X + \frac{N^2}{288} \Tr(\cR^2)+\frac{M}{24} X^T \cM^{(0)} \cR X- \frac{N}{24}  X^T \Omega \cR X \nnr
&+\frac{1}{16}\mathcal I_4(X)+ \frac{1}{24}T_4 - \frac{1}{192} (X^T \mathcal M^{(0)} X)^2 - \frac{1}{2^{10}3^4}\Tr(\cR^2)^2 \nnr
& +  \frac{5}{2^{11}3^4}  \sqrt{\Tr(\cR^2)^4 - (2^8 3^3 5^{-3} 11) \Tr(\cR^2) \Tr(\cR^6) + (2^9 3^6 5^{-3}) \Tr(\cR^8)}.
\label{FinvE7}
\end{align}
As was emphasized before, the entropy of the STU black hole will not change under $E_{7(7)}$ dualities (since the four-dimensional metric is invariant) and the STU black hole is a seed for the generic $\mathcal N = 8$ black hole. Therefore, although \eqref{FinvE7} was derived in the STU truncation, it is the correct formula for the most general non-extremal black hole of $\mathcal N = 8$ supergravity. The entropy of such a black hole takes the form \eqref{NLR}, where the quartic invariant $\Delta$ is given by \eqref{I4} and \eqref{eq:I4E7} and the F-invariant is given by \eqref{FinvE7}\footnote{Our final formula differs from \cite{Sarosi:2015nja} as discussed in the summary.}.

\section{$\mathcal N = 2$ supergravities and dualities}
\label{sec:N2}

Let us now turn to the formalism of ungauged four-dimensional $\mathcal N=2$ supergravity coupled to $n_v$ gauge multiplets without hypermultiplets. The theory admits $n_v$ complex scalars $z_i$, $i=1,\dots,n_v$, which define a special K\"alher manifold. The special geometry of the scalar manifold is naturally encoded in $U(1) \times Sp(2n_v+2,\R)$ symplectic sections 
\be
\mathcal V = \left(X^\Lambda,F_\Lambda \right),\qquad \Lambda=0, \dots , \, n_v,
\ee
where $F(X)$ is the prepotential (a homogeneous meromorphic function of degree 2) and $F_\Lambda \equiv \frac{\p}{\p X^\Lambda}F$. We are mainly interested in cubic prepotentials which have the form 
\be
F = -\frac{1}{6} C_{ijk} \frac{X^i X^j X^k}{X^0} \equiv - \frac{N_3[X]}{X^0} 
\ee 
where we introduced the cubic norm $N_3[X]$. Here $C_{ijk}$ is completely symmetric.  The $n_V+1$ complex fields are constrained by $\text{Im} F_{\Lambda M} X^\Lambda \bar X^M =-1$ %-\frac{1}{16\pi G}$ 
where $F_{\Lambda M}=\p_{X^\Lambda}\p_{X^M}F$. The resolution of the constraints determines the scalar fields $z_i$ in a fixed $U(1)$ gauge.  The K\"ahler potential is then defined up to an arbitrary holomorphic function as $\mathcal K = -\log (- i N[z-\bar z])$. 
The STU model admits $n_v=3$ gauge multiplets and the prepotential $F(X) = -X^1 X^2 X^3/X^0$. The scalar fields are then defined as $z^i = X^i/X^0$. 
In the following, we will keep the cubic prepotential arbitary. We will also comment at the end of the section on the F-invariant expression for quadratic prepotentials. 

\subsection{Covariant tensors}

In $\cN = 2$ formalism, the charges belong to a symplectic vector
\be \Gamma = \left( \begin{array}{c} \wtd{P}_\Lambda \\ \wtd{Q}^\Lambda \end{array} \right)  , \ee
with $\Lambda=0,\dots n_v$. The charge vector transforms under an element $g \in Sp(2n_v+2,\R)$ as
\be \Gamma \mapsto g \Gamma 
\ee
where 
\be
g^T \Omega g = \Omega, \qquad\quad \Omega= \begin{pmatrix} 0 & I_{(n_v+1)\times (n_v+1)} \\ -I_{(n_v+1)\times (n_v+1)} & 0 \end{pmatrix}. \ee
For the case of the STU model and in the notations of the previous section, the tilded charges are related to the untilded charges by the relations \cite{Chow:2014cca}
\be \wtd{P}_0=P^4, \quad \wtd{P}_i = - Q_i, \quad \wtd{Q}^0 = Q_4, \quad \wtd{Q}^i = P^i \qquad (i=1,2,3) .\ee
These relations are also consistent with \eqref{tQP}. 
The transformation \eqref{SL2charges} of the charges defines a matrix $g \in Sp(8,\R)$, which realizes the group embedding $SL(2,\R)^3 \subset Sp(8,\R)$.

The standard definitions of central charge and K\"ahler metric are 
\bea
Z[z,\bar z, \wtd{P}, \wtd{Q}] = e^{\frac{\mathcal K}{2}}\mathcal V\, \Omega \, \Gamma,\qquad g_{i \bar{j}}[z,\bar z] = \p_i \p_{\bar{j}}\mathcal K(z,\bar z). 
\eea
We also define the K\"alher derivative of the central charge as $Z_i = D_i Z \equiv (\p_{z_i} + \frac{1}{2}\p_{z_i}\mathcal K)Z$. We can already define the following invariants \cite{Ceresole:2009iy}: 
\begin{align}
i_1 &= Z \bar Z , & i_2 &= g^{i \bar{j} }Z_i \bar Z_{\bar j},\label{defi1i2}\\
i_3 &= \text{Re} (Z N_3[\bar Z]), & i_4 &= - \text{Im} (Z N_3[\bar Z]),\\
i_5 &= g^{i \bar i}C_{ijk}C_{\bar i \bar j \bar k} Z^j  Z^k \bar Z^{\bar j} \bar Z^{\bar k}.
\end{align}
The last three invariants are only defined for cubic prepotentials while the first two are defined for arbitrary prepotentials. 
The quartic invariant is given by
\bea
16\Delta = (i_1-i_2)^2 + 4 i_4 - i_5. 
\eea
It is independent of the scalar fields. At leading order in the radial asymptotic expansion, the scalar fields take their moduli values and therefore $i_1,i_2$ and $i_3,i_4$ at radial infinity, which we denote as $i^\infty_1,i^\infty_2$ and $i^\infty_3,i^\infty_4$ provide independent invariants. 

Let us now define other invariants. In general, the period matrix is defined from the scalars as 
\bea
\cN_{\Lambda M} = \bar F_{\Lambda M}+2 i \frac{F_{\Lambda N} X^N F_{M \Xi }X^\Xi }{ X^O F_{O \Pi} X^\Pi}
\eea
where $F_{\Lambda M}=\p_{X^\Lambda}\p_{X^M}F$. 
Then, the $2(n_v+1)\times 2(n_v+1)$ matrix
\be \wh{\cN} =
\begin{pmatrix}
\Imag \mathcal N + \Real \mathcal N (\Imag \mathcal N)^{-1} \Real \mathcal N  & - \Real \mathcal N (\Imag \mathcal N)^{-1} \\
-(\Imag \mathcal N)^{-1} \Real\mathcal N & (\Imag \mathcal N)^{-1}
\end{pmatrix}
\ee
transforms as
\be \wh{\cN} \mapsto (g^{-1})^T \wh{\cN} g^{-1} . \ee
In the example of the STU model, we find the following $4 \times 4$ complex matrix after fixing the $U(1)$ gauge to $X^0 =1$, 
\be
\cN =
\begin{pmatrix}
-2 x_1 x_2 x_3 - \im y_1 y_2 y_3 \big( 1 + \sum_{i=1}^3 \frac{x_i^2}{y_i^2} \big) & x_2 x_3 + \im \frac{x_1 y_2 y_3}{y_1} & x_1 x_3 + \im \frac{x_2 y_1 y_3}{y_2} & x_1 x_2 + \im \frac{x_3 y_1 y_2}{y_3} \\
%%%
x_2 x_3 + \im \frac{x_1 y_2 y_3}{y_1} & - \im \frac{y_2 y_3}{y_1} & - x_3 & -x_2 \\
%%%
x_1 x_3 + \im \frac{x_2 y_1 y_3}{y_2} & - x_3 & - \im \frac{y_1 y_3}{y_2} & - x_1 \\
%%%
x_1 x_2 + \im \frac{x_3 y_1 y_2}{y_3} & -x_2 & -x_1 & -\im \frac{y_1 y_2}{y_3} \end{pmatrix}
.
\ee
We assume the radial asymptotic expansion
\be \wh{\cN} = \wh{\cN}^{(0)} + \frac{\wh{\cN}^{(1)}}{r} +  \mathcal{O}\left(\frac{1}{r^2}\right), \ee
which is clearly correct for the STU model. It is then natural to define the matrix
\be \wh{R} = (\wh{\cN}^{(0)})^{-1} \wh{\cN}^{(1)}, \ee
that transforms as
\be \wh{R} \mapsto g \wh{R} g^{-1}. \ee

One usually defines the $2(n_v+1)\times 2(n_v+1)$ matrix
\be \wh{\cF} =
\begin{pmatrix}
\Imag \mathcal F + \Real \mathcal F(\Imag \mathcal F)^{-1} \Real \mathcal F  & - \Real \mathcal F (\Imag \mathcal F)^{-1} \\
-(\Imag \mathcal F)^{-1} \Real\mathcal F & (\Imag \mathcal F)^{-1}
\end{pmatrix},\label{defF}
\ee
where $\cF = (F_{\Lambda M})$. It also transforms as
\be \wh{\cF} \mapsto (g^{-1})^T \wh{\cF} g^{-1} . \ee
We will use this quantity in the case of quadratic prepotentials.

\subsection{Construction of invariants}

The list of relevant invariants for generic cubic $\mathcal N = 2$ supergravities is given by
\begin{itemize}
\item Degree 1:
\begin{align}
M,\, N.
\end{align}
\item Degree 2: 
\begin{align}
  \Gamma^T \wh{\cN}^{(0)} \Gamma = 2(i_1^\infty + i_2^\infty) ,\qquad \Tr(\wh{R}^2).
\end{align}
\item Degree 3:
\begin{align}
\Gamma^T \Omega \wh{R} \Gamma,\qquad \Gamma^T \wh{\cN}^{(0)} \wh{R} \Gamma .
\end{align}
\item Degree 4:
\begin{align}
\Delta ,\qquad i^\infty_4 ,\qquad  \Tr(\wh{R}^4).
\end{align}
\end{itemize}
In the case of the STU model, it is straightforward to check that these invariants relate to the ones defined in Section \ref{consSTU} as
\begin{itemize}
\item Degree 2: 
\begin{align}
L_1 & = - \Gamma^T \wh{\cN}^{(0)} \Gamma = -2 (i_1^\infty + i_2^\infty), \label{relL1i1i2}\\
L_2 &= \frac{1}{12} \Tr(\wh{R}^2).
\end{align}
\item Degree 3:
\begin{align}
C_1 &= \frac{1}{3}\Gamma^T \Omega \wh{R} \Gamma,\\
 C_2 &= \frac{1}{3}\Gamma^T \wh{\cN}^{(0)} \wh{R} \Gamma .
\end{align}
\item Degree 4:
\begin{align}
 \Delta_2 &= \frac{1}{3} (\Delta- i^\infty_4 ) + \frac{1}{96} (\Gamma^T \wh{\cN}^{(0)} \Gamma)^2 ,\\
 32 \Delta_3 &= -\frac{1}{24} \Tr(\wh{R}^4) + \frac{1}{64} \Tr(\wh{R}^2)^2.
\end{align}
\end{itemize}

\subsection{The F-invariant}

Given the dictionary above, one can reexpress the formula for the F-invariant given in \eqref{Finv1} in terms of invariants specific to $\mathcal N=2$ supergravities with cubic prepotential
as follows
\begin{align}\label{Finv2}
F = &M^4 + M^2 N^2 + \frac{1}{4} M^2 \Gamma^T \wh{\cN}^{(0)} \Gamma + \frac{1}{96} N^2 \Tr(\wh{R}^2) + \frac{1}{24} M \Gamma^T \wh{\cN}^{(0)} \wh{R} \Gamma - \frac{1}{24} N \Gamma^T \Omega \wh{R} \Gamma \nonumber\\
&- \frac{1}{192} \left( 64  \Delta + 32 i^\infty_4 + \frac{1}{8} \Tr(\wh{R}^4) - \left(\Gamma^T \wh{\cN}^{(0)}\Gamma\right)^2 - \frac{1}{64} \Tr(\wh{R}^2)^2 \right).
\end{align}
Since this formula makes sense for arbitrary cubic prepotential, it is a natural ansatz for the F-invariant in the generic case. However, this would need to be checked with explicit non-extremal black hole solutions different than STU black holes, see e.g. \cite{Kastor:1997wi}.

Let us finally comment on  quadratic prepotentials. We will only discuss the case of the $\overline{\mathbb{CP}}^n$ model with prepotential
\bea
F=-\frac{i}{4}\eta_{\Lambda \Sigma}X^\Lambda X^\Sigma, \qquad \eta_{\Lambda \Sigma} = \text{diag}(+-\dots -).
\eea
A general class of non-extremal dyonic rotating black holes with moduli without NUT charge for this model was found in \cite{Galli:2011fq}. It was noted there that the black hole entropy $S_+$ and the fake ``inner horizon entropy'' $S_-$, defined as the area of the inner horizon divided by $4G$, can be written as 
\bea
S_\pm  = \pi \left( \sqrt{\hat N_R} \pm \sqrt{\hat N_L} \right)^2,\qquad \hat N_{R,L} =M^2 - \frac{1}{2}(|Z_\infty|^2+ |D_i Z_\infty|^2 ) \pm \sqrt{\widehat \Delta + J^2}
\eea
where $\widehat \Delta = \frac{1}{4}(|Z_\infty|^2- |D_i Z_\infty|^2)^2$ is a square of quadratic invariants and $J$ the angular momentum. Here, one might substitute $|Z_\infty|^2+ |D_i Z_\infty|^2 = i_1^\infty + i_2^\infty = \frac{1}{2}  \Gamma^T \wh{\cN}^{(0)} \Gamma$ and $\widehat \Delta = \frac{1}{4}(i_1^\infty - i_2^\infty)^2$ using \eqref{defi1i2} and \eqref{relL1i1i2}. We also observe that $\widehat \Delta =\frac{1}{16}  ( \Gamma^T \wh{\cF}^{(0)} \Gamma )^2 $ where $\wh{\cF}^{(0)} $ is defined as the leading term in the radial expansion of $\wh \cF$ defined in \eqref{defF}.  Now we simply note that these formulas can also be written in the Cardy form
\bea
S_\pm  =2 \pi \left( \sqrt{N_L} \pm \sqrt{N_R} \right),\qquad N_L=F+\widehat \Delta,\qquad N_R = F-J^2,
\eea
where the F-invariant takes the manifestly duality invariant form
\bea
F = \left(M^2- \frac{1}{4}  \Gamma^T \wh{\cN}^{(0)} \Gamma\right)^2 - \widehat \Delta. 
\eea
It is natural to expect that other quadratic models will admit the same entropy formula. 

%% OLD: S_1=32\Delta ; S_2 = 32\Delta_2 ; S_3 = 32 \Delta_3 ; I_4 = 16 \Delta

\appendix
\section{General non-extremal black hole with moduli}
\label{app1}

Non-trivial scalar moduli at infinity are generated by the $SL(2,\R)^3$ transformations
\be S_i = \begin{pmatrix} e^{-\varphi_i^\infty/2} & \chi_i^\infty e^{\varphi_i^\infty/2} \\ 0 & e^{\varphi_i^\infty/2} \end{pmatrix}, \ee
which act on the scalars as
\be \varphi_i \mapsto \varphi_i + \varphi_i^\infty, \qquad \chi_i \mapsto \chi_i e^{-\varphi_i^\infty} + \chi_i^\infty .\ee
The action on the charges is given by
\be \begin{pmatrix} P^I \\ -Q_I \end{pmatrix} \mapsto \mathbb{S} \begin{pmatrix} P^I \\ -Q_I \end{pmatrix}, \ee
where $\mathbb{S}$ is the moduli-dependent $8\times 8$ symplectic matrix $\mathbb{S} =  e^{-(\vph_1^\infty+\vph_2^\infty+\vph_3^\infty)/2} \mathbb{T}$, 
\be
\mathbb{T}=\begin{pmatrix}
e^{\vph_1^\infty} & 0 & 0 &-\lambda_2\lambda_3e^{\vph_1^\infty}&0&-\lambda_3e^{\vph_1^\infty}&-\lambda_2e^{\vph_1^\infty}&0 \\
0 & e^{\vph_2^\infty}&0&-\lambda_1\lambda_3e^{\vph_2^\infty}&-\lambda_3e^{\vph_2^\infty}&0&-\lambda_1e^{\vph_2^\infty}&0 \\
0&0&e^{\vph_3^\infty}&-\lambda_1\lambda_2e^{\vph_3^\infty}&-\lambda_2e^{\vph_3^\infty}&-\lambda_3e^{\vph_3^\infty}&0&0 \\
0&0&0&e^{\vph_1^\infty+\vph_2^\infty+\vph_3^\infty}&0&0&0&0 \\
0&0&0&\lambda_1e^{\vph_2^\infty+\vph_3^\infty}&e^{\vph_2^\infty+\vph_3^\infty}&0&0&0\\
0&0&0&\lambda_2e^{\vph_1^\infty+\vph_3^\infty}&0&e^{\vph_1^\infty+\vph_3^\infty}&0&0\\
0&0&0&\lambda_3e^{\vph_1^\infty+\vph_2^\infty}&0&0&e^{\vph_1^\infty+\vph_2^\infty}&0\\
\lambda_1&\lambda_2&\lambda_3 & -\lambda_1\lambda_2\lambda_3&-\lambda_2\lambda_3&-\lambda_1\lambda_3&-\lambda_1\lambda_2&1
\end{pmatrix}
\ee
and $\lambda_i = \chi_i^\infty e^{\vph_i^\infty}$.

This can be used to generalize the non-extremal black hole solution of \cite{Chow:2014cca} to non-trivial scalar moduli at infinity. Defining the barred quantities to be those of \cite{Chow:2014cca}, the generalized solution is\footnote{In the $\mathfrak{so}(4,4)$ formalism of \cite{Chow:2014cca}, this corresponds to acting on the 3d coset matrix of the solution with trivial moduli with
\[ g_\text{mod} = \exp\left(\frac{1}{2}\vph_1^\infty H_1 + \frac{1}{2}\vph_2^\infty H_2 + \frac{1}{2}\vph_3^\infty H_3 \right) \exp\left( -\chi_1^\infty E_1 - \chi_2^\infty E_2 - \chi_3^\infty E_3 \right), \]
where $H_i$, $E_i$ are the $\mathfrak{so}(4,4)$ generators defined in \cite{Chow:2014cca}.}
\begin{align}
ds^2 &= \overline{ds^2} ,\\
\vph_i &= \overline{\vph_i} + \vph_i^\infty, \\
\chi_i &= e^{-\vph_i^\infty} \overline{\chi_i} + \chi_i^\infty ,\\
\begin{pmatrix} A^I \\ \widetilde{A}_I \end{pmatrix} &=  \mathbb{S} \,\begin{pmatrix} \overline{A}^I \\ \overline{\widetilde{A}\hspace{2pt}}_I .\end{pmatrix} .
\end{align}
It depends upon 17 parameters: the mass parameter $m$, the NUT parameter $n$, the rotation parameter $a$, the 8 charging parameters $\delta_I$ and $\gamma_I$, and the 6 moduli parameters $\vph_i^\infty$ and $\chi_i^\infty$. These parameters can in principle be traded for the physical charges, but it is algebraically very complicated in general.

\section{$\mathfrak{so}(8)$ generators $\Gamma^{ij}$}
\label{app:so8}

The matrices $\Gamma^{ij}$ ($i,j=1, \dotsc, 8$) that we used to change basis from $\mathfrak{sl}(8)$ to $\mathfrak{su}(8)$ in the $\mathfrak{e}_{7(7)}$ algebra are given by
\begin{align}
\Gamma^{ij} &= \gamma^{[i} \gamma^{j]} ,\nnr
\Gamma^{i 8} &= - \gamma^i \qquad (\Gamma^{8 i} = \gamma^i) ,
\end{align}
($i,j=1, \dotsc, 7$), and the $\gamma^i$ are $\mathfrak{so}(7)$ gamma matrices satisfying
\be \{ \gamma^i, \gamma^j \} = - 2 \delta^{ij} I_{8\times 8} . \ee
It follows from this definition and the Clifford algebra of the $\gamma^i$ that the $\Sigma^{ij}=-\frac{1}{2}\Gamma^{ij}$ satisfy the $\mathfrak{so}(8)$ commutation relations
\be \left[ \Sigma^{ij},\Sigma^{kl}\right] = \delta^{il} \Sigma^{jk} + \delta^{jk} \Sigma^{il} - \delta^{ik} \Sigma^{jl} - \delta^{jl} \Sigma^{ik}. \ee

For explicit computations, we used the real antisymmetric representation
\begin{align}
\gamma^1 &= -i \sigma_3 \otimes \sigma_2 \otimes \sigma_1, \nnr
\gamma^2 &= i \sigma_3 \otimes \sigma_2 \otimes \sigma_3, \nnr
\gamma^3 &= -i \sigma_3 \otimes I_{2\times 2} \otimes \sigma_2 ,\nnr
\gamma^4 &= -i \sigma_1 \otimes \sigma_1 \otimes \sigma_2, \nnr
\gamma^5 &= -i \sigma_1 \otimes \sigma_2 \otimes I_{2\times 2} ,\nnr
\gamma^6 &= i \sigma_1 \otimes \sigma_3 \otimes \sigma_2 ,\nnr
\gamma^7 &= i \sigma_2 \otimes I_{2\times 2} \otimes I_{2\times 2}
\end{align}
where the $\sigma_i$ are the standard Pauli matrices.

\section*{Acknowledgments}

The authors would like to thank M. Henneaux for useful conversations and D. Chow, M. Esole and G. S\'arosi for an interesting correspondence.  G.C. is Research Associate of the Fonds de la Recherche Scientifique F.R.S.-FNRS (Belgium) and acknowledges the current support of the ERC Starting Grant 335146 ``HoloBHC". V.L. is Research Fellow of the Fonds de la Recherche Scientifique F.R.S.-FNRS (Belgium) and was partially
funded by the ERC through the SyDuGraM Advanced Grant. This work is also partially supported by FNRS-Belgium (convention IISN 4.4503.15).

%\bibliography{refs}

\providecommand{\href}[2]{#2}\begingroup\raggedright\endgroup

\end{document}